\documentclass[12pt,preprint]{aastex}
\usepackage{natbib}
\usepackage{graphicx}
\usepackage{epsfig}

\newcommand{\fesc}{\ifmmode{f_{\rm esc}}\else{$f_{\rm esc}$}\fi}
\newcommand{\fescs}{\ifmmode{f_{\rm esc}^\star}\else{$f_{\rm esc}^\star$}\fi}
\newcommand{\kms}{\ifmmode{{\rm km~s^{-1} }}\else{km s$^{-1}$}\fi}
\newcommand{\cubecm}{\ifmmode{{\rm cm^{-3} }}\else{cm$^{-3}$}\fi}
\newcommand{\lsim}{\lower0.3em\hbox{$\,\buildrel <\over\sim\,$}}
\newcommand{\gsim}{\lower0.3em\hbox{$\,\buildrel >\over\sim\,$}}

\newcommand{\Ms}{\ifmmode{M_\odot}\else{$M_\odot$}\fi}

\newcommand{\mvir}{\ifmmode{M_{\rm{vir} }}\else{$M_{\rm{vir}}$}\fi}
\newcommand{\rvir}{\ifmmode{r_{\rm{vir} }}\else{$r_{\rm{vir}}$}\fi}

\newcommand{\jj}{\ifmmode{J_{21}}\else{$J_{21}$}\fi}
\newcommand{\flw}{\ifmmode{F_{LW}}\else{$F_{LW}$}\fi}

\begin{document}

\shorttitle{}
\shortauthors{}

\title{A General Formula for Black Hole Gravitational Wave Kicks}

\author{
James~R. van~Meter \altaffilmark{1,2}, 
M.~Coleman Miller \altaffilmark{3,4}, 
John G. Baker \altaffilmark{5}, 
William~D.~Boggs \altaffilmark{1,6}, 
Bernard~J.~Kelly \altaffilmark{1,2}
}

\altaffiltext{1}{CRESST and Gravitational Astrophysics Laboratory, NASA/GSFC, Greenbelt, MD 20771}
\altaffiltext{2}{Department of Physics, University of Maryland, Baltimore County, 1000 Hilltop Circle, Baltimore, MD 21250}
\altaffiltext{3}{Department of Astronomy, University of Maryland, College Park, MD 20742}
\altaffiltext{4}{Joint Space Science Institute, University of Maryland, College Park, MD 20742}
\altaffiltext{5}{Gravitational Astrophysics Laboratory, NASA/GSFC, Greenbelt, MD 20771}
\altaffiltext{6}{Department of Physics, University of Maryland, College Park, MD 20742}

\email{james.r.vanmeter@nasa.gov}

\begin{abstract}

Although the gravitational wave kick velocity in the orbital plane
of coalescing black holes has been understood for some time, 
apparently conflicting formulae have been proposed for the
dominant  out-of-plane kick, each a good fit to different data
sets. This is important to resolve because it is only the
out-of-plane kicks that can reach more than 500~\kms and
can thus eject merged remnants from galaxies.  Using a different
ansatz for the out-of-plane kick, we show that we can fit almost
all existing data to better than 5\%.  This is good enough for any
astrophysical calculation, and shows that the previous apparent
conflict was only because the two data sets explored different
aspects of the kick parameter space.
  
\end{abstract}


\section{Introduction}

When two black holes spiral together and merge, the gravitational
radiation they emit is usually asymmetric and thus the remnant black
hole acquires a linear velocity relative to the original center of
mass.  The speed can reach over 3,000~\kms 
\citep{Dain:2008ck}, which
would eject the remnant from any galaxy in the universe
(see Figure~2 of \citet{Merritt:2004xa}).  As discussed in
\citet{Merritt:2004xa}, the kick magnitude and distribution
are important for discussions of hierarchical
merging, supermassive black hole formation, galactic nuclear dynamics,
and the degree to which black holes influence galaxy formation.  
The community has converged on the
formula for the kick speed in the original orbital plane 
\citep{Baker:2007gi,Campanelli:2007cg,Gonzalez:2006md}, but
apparently conflicting dependences for the out-of-plane kick (which
dominates the total kick for most configurations) have been proposed.
\cite{Lousto:2008dn} suggested that for a binary with component masses
$m_1$ and $m_2\geq m_1$, the kick scales as $\eta^2$, where $\eta\equiv
m_1m_2/(m_1+m_2)^2$ is the \emph{symmetric mass ratio}; however the data in
\cite{Baker:2008md} were fit much better with
an $\eta^3$ dependence.  The conflict is only apparent, however,
because there were no runs in common between the two data sets.
This suggests that an analysis might be performed with a new ansatz that
can fit all of the existing data.

Here we perform such a fit, and demonstrate that there is a single
formula for the out-of-plane kick that fits almost all existing data
to better than 5\% accuracy.  
The new ansatz is similar to one recently suggested by \cite{Lousto:2009mf}, for which, however, 
a fit was not attempted.  Its form is based straightforwardly on the post-Newtonian (PN)
approximation and includes both the 
aforementioned $\eta^2$ and $\eta^3$ terms, as well as a slightly more complicated spin-angle dependence.
In \S~2 we describe our new runs and
we describe the new ansatz in \S~3.  In \S~4 we list all 95 runs we have
fit, from different numerical relativity groups, and our fitting
procedure and best-fit parameters.  In \S~5 we discuss the
implications of our results, and indicate the fraction of kicks
above 500~\kms and 1,000~\kms for representative spin
and mass ratio distributions, comparing it with previous results.
The goodness of our fit to the entire usable data set of
out-of-plane kicks suggests that the full three-dimensional kick is
now modelled well enough that it will not limit the accuracy of any
astrophysical calculation.

\section{Numerical simulations}
\label{sec:simulations}

We have performed new simulations representing 22 distinct
physical cases.   Defining $q\equiv m_1/m_2$ and $\alpha_i\equiv
S_i/m_i^2$ where  $S_i$ is the spin angular momentum of the $i$th
black hole, we used mass ratios of $q=0.674$ or $q=0.515$,  and
spins initially within the orbital plane of $\alpha_1=\alpha_1^{\perp}=0.367$ and
$\alpha_2=\alpha_1^{\perp}=0.177$ or $\alpha_2=\alpha_1^{\perp}=0.236$, where the
$\perp$ superscript indicates orthogonality to the orbital axis. 
For each mass ratio we used
one of eleven different spin orientations (given in Table~1), in order to probe the
spin-angle dependence of the final recoil.   Initial parameters were
informed by a quasicircular PN approximation  and
initial data constructed using the spectral solver {\tt
TwoPunctures} \citep{Ansorg:2004ds}. Evolutions were performed with
the Einstein-solver {\tt Hahndol}
\citep{Imbiriba:2004tp,vanMeter:2006vi,Baker:2007gi}, with all
finite differencing and interpolation at least fifth-order-accurate
in computational grid spacing.  Note for the purpose of
characterizing the simulations, it is convenient to use units in
which  $G=c=1$ and specify distance and time in terms of $M\equiv
m_1+m_2$. The radiation field represented by the Weyl scalar
$\Psi_4$ was interpolated to an extraction sphere  at coordinate
radius $r=50M$ and integrated to obtain the radiated momentum using
the standard formula \citep{Schnittman:2007ij}:
\begin{equation}
P_i=\int_{-\infty}^t dt\frac{r^2}{16\pi}\int d\Omega \frac{x_i}{r}\left|\int_{-\infty}^t dt\Psi_4 \right|^2.
\end{equation}
For each of the 22
cases we ran two resolutions, with fine-grid spacings of $h_f=3M/160$
and $h_f=M/64$.  
To give some indication of our numerical error, 
the total recoil from the two resolutions for each of
the $q=0.674$ cases agreed to within $\lsim 6\%$, and for each of the
$q=0.515$ cases to within $1\%$ (the latter using a more optimal grid structure).

In these simulations, variation in the magnitude of the final recoil within the $x$-$y$ plane
suggested non-negligible precession of the orbital plane.  Indeed the coordinate trajectories
of the black holes showed precession of up to $\sim 10^{\circ}$.  
To calculate the component of the recoil velocity parallel to the ``final" orbital axis, $V_{||}$, we tried
two different methods. 
In one method we simply took the dot product of the final, numerically computed recoil velocity ${\mathbf V}$ with the 
normalized orbital angular momentum ${\mathbf L}|L|^{-1}$ as calculated from the coordinate trajectories of the black holes,
just when the common apparent horizon was found:
$V_{||}\equiv {\mathbf V}\cdot{\mathbf L }{|L|}^{-1}$.
In our second method we assumed that each black hole spin, which is
initially orthogonal to the orbital angular momentum,
remained approximately orthogonal throughout the simulation, i.e. 
\begin{equation}
\label{eq:LS0}
{\mathbf L}\cdot{\mathbf S}_1\approx{\mathbf L}\cdot{\mathbf S}_2\approx 0.
\end{equation}
This assumption is consistent with PN calculations, to linear order in spin \citep{Racine:2008qv,Kidder:1995zr}, 
and is also supported numerically by the fact that the merger times in our simulations were
independent of the initial spin, to within $\Delta t<1M$.  
In this case the in-plane recoil should depend
only on the mass ratio (and not the spin), and we assume it is given by the formula found by \cite{Gonzalez:2006md},
with coefficient values given by a previous fit \citep{Baker:2008md}.  This implies
\begin{equation}
\label{eq:vpar}
V_{||}\equiv \sqrt{V^2-(V_{\perp}^{pred})^2}\frac{V_z}{|V_z|},
\end{equation} 
where we have further assumed that the sign of $V_{||}$ should be the same as that of $V_z$, given the 
modest amount of precession.  

These two definitions for $V_{||}$  were found to differ by a
relative error of $\lsim 5\%$ for all points except one for which
$V_{||} \ll V_{\perp}$.  Relative to the maximum $V_{||}$ per mass
ratio, they were found to differ only by $<2\%$.
Note this lends further support to our assumption that the spins are
orthogonal to the orbital angular momentum to good approximation
throughout these simulations (Eq.~(\ref{eq:LS0})).
Even $V_z$ was
found to differ from the above definitions for $V_{||}$  by only
$\lsim 2\%$, relative to the maximum.  We will use
Eq.~(\ref{eq:vpar}) to give our canonical $V_{||}$, for the purpose
of analytic fits to the data.

An additional assumption we will make about our data,
important for fitting purposes, 
is that the amount of precession undergone by the spins in the
orbital plane, i.e. the difference in spin-angles, 
between the initial data and the merger, is independent
of the initial spin orientation. This assumption, valid to linear
order in spin according to the PN approximation (Eq.~(2.4) of \citep{Kidder:1995zr}), was
previously found to be the case through explicit computation of the
spins in the simulations presented in \cite{Baker:2008md}.
We have not explicitly computed the spins in the new simulations
presented here but the independence of the merger time with respect to
initial spin orientation is consistent with the assumption of similar
independence of spin precession.

For the purpose of constructing an accurate phenomenological model, 
we added to this data set previously published data representing
out-of-plane kicks.  Our criteria for selecting relevant data are
that  the component of the final recoil parallel to the orbital plane
is given and at least three different in-plane spin orientations are
used.  We arrived at a total of 95 data points: the 22 new ones presented
here plus 73 drawn from 
\cite{Baker:2008md,Campanelli:2007cg,Dain:2008ck,Lousto:2008dn}.
Note that for the cases in \cite{Lousto:2008dn} we use the second version
of their values listed in their Table III, resulting from what they
consider to be their best calculation of the final orbital plane.

\section{Ansatz}

An ansatz for the total recoil that was found to be very consistent with
numerical results has the form \citep{Baker:2007gi,Campanelli:2007cg,Gonzalez:2006md}
\begin{eqnarray}
\vec{V} &=& V_{\perp m} \, {\mathbf e}_1 + V_{\perp s} (\cos\xi \, {\mathbf e}_1 + \sin\xi \, {\mathbf e}_2) + V_{\parallel} \, {\mathbf e}_3, \label{eq:V_total}\\
      V_{\perp m}    &=& A \eta^2 \sqrt{1 - 4 \eta} (1 + B \eta), \label{eq:V_mass_Gonzalez}\\
V_{\perp s}     &=& H \frac{\eta^2}{(1+q)} \left( \alpha_2^{\parallel} - q \alpha_1^{\parallel} \right), \label{eq:V_perp}
\end{eqnarray}
where ${\mathbf e}_1$ and ${\mathbf e}_3$ are unit vectors in the directions of separation and the orbital axis just before merger, respectively,
${\mathbf e}_2\equiv{\mathbf e}_1\times{\mathbf e}_3$, $\alpha_1^{\parallel}$ and $\alpha_2^{\parallel}$ represent the components of spins parallel to the orbital axis,
$\xi, A, B$, and $H$ are constant fitting parameters, and $V_{\parallel}$ is to be discussed.

Every component of Eq.~(\ref{eq:V_total}) has been modeled to some degree after PN expressions.
The use of such PN-based formulae has proven very successful.  
For example, Eq.~(\ref{eq:V_mass_Gonzalez}) for the in-plane, mass-ratio-determined
recoil can be obtained from the corresponding PN expression (Eq.~(23) of \cite{Blanchet:2005rj})
simply by taking the leading-order terms and replacing
instances of the PN expansion parameter (in this case frequency)
with constant fitting parameters.
Using this, \cite{Gonzalez:2006md} obtained very good agreement with a large
set of numerical results.
\cite{Tiec:2009yg}, who calculated the recoil using a combination
of the PN method with a perturbative approximation of the ringdown,
also found that Eq.~(\ref{eq:V_mass_Gonzalez}) gave a good phenomenological fit to their analytic results.

Why such a prescription for generating an ansatz 
from the PN approximation should apply so well to merger dynamics is not perfectly understood.
The effective replacement of powers of the frequency with constants
may be defensable because the majority of the recoil is generated within a narrow time-window near merger,
perhaps within a narrow range of frequencies. 
This is particularly evident for the out-of-plane recoil speed,
which rapidly and monotonically increases up to a constant value
around merger.
However, values of the constant coefficients that appear in the PN expansion 
cannot be expected to remain unchanged because, as merger is approached, high-order PN terms
with the same functional dependence on mass and spin as leading order
terms can become comparable in magnitude.  So, it is just for
this functional dependence that we look to the PN approximation
for guidance.

Following the prescription implied above, from the PN expression
for out-of-plane recoil given by \cite{Racine:2008kj},
Eqs.~(4.40-4.42) (neglecting spin-spin interaction),
the following can be straightforwardly obtained:
\begin{eqnarray}
V_{||} &=& \frac{K_2\eta^2+K_3\eta^3}{q+1}\left[q \alpha^{\perp}_1
\cos(\phi_1-\Phi_1)-\alpha^{\perp}_2\cos(\phi_2-\Phi_2)\right] \nonumber\\
  &+&\frac{K_S(q-1)\eta^2}{(q+1)^3}\left[q^2\alpha^{\perp}_1
\cos(\phi_1-\Phi_1)+\alpha^{\perp}_2\cos(\phi_2-\Phi_2)\right],
\end{eqnarray}
where $K_2$, $K_3$, and $K_S$ are constants, $\alpha^{\perp}_i$ represents the magnitude of the projection of the $i$th black hole's spin (divided by the square of
the black hole's mass) into the orbital plane, $\phi_i$ represents the angle of the same projection,
as measured at some point before merger, with respect to a reference angle representing the direction of separation of the black holes, and $\Phi_i$ represents
the amount by which this angle precesses before merger, which depends on the mass ratio and the initial separation.   
We have ignored terms quadratic in the spin because we assume they are subleading.
Note that this ansatz is equivalent to one suggested by \cite{Lousto:2009mf}, provided the angular parameters are suitably interpreted.
In terms of the notation of \cite{Racine:2008kj},
\begin{eqnarray}
(q+1)^{-1}\left[q \alpha^{\perp}_1 \cos(\phi_1-\Phi_1)-\alpha^{\perp}_2\cos(\phi_2-\Phi_2)\right]   =& -M^{-1}{\mathbf \Delta}\cdot\mathbf{n} &=  -M^{-1}\Delta^{\perp}\cos(\Theta), \\
(q+1)^{-2}\left[q^2\alpha^{\perp}_1 \cos(\phi_1-\Phi_1)+\alpha^{\perp}_2\cos(\phi_2-\Phi_2)\right]  =&  M^{-2}{\mathbf S}\cdot\mathbf{n}      &=  M^{-2}S^{\perp}\cos({\Psi}),
\end{eqnarray}
where ${\mathbf{n}}$ is a unit separation vector, $\Delta\equiv \mathbf{S}_2/m_2-\mathbf{S}_1/m_1$,
$\mathbf{S}\equiv \mathbf{S}_1+\mathbf{S}_2$,
$\Delta^{\perp}\equiv M (q+1)^{-1}|\alpha^{\perp}_2-q\alpha^{\perp}_1|$,
$S^{\perp}\equiv M^2 (q+1)^{-2}|\alpha^{\perp}_2+q^2\alpha^{\perp}_1|$, 
$\Theta$ is the angle between $\Delta$ and ${\mathbf{n}}$, 
and $\Psi$ is the angle between $\mathbf{S}$ and ${\mathbf{n}}$,
all measured, for our purposes, at some point arbitrarily close to merger.
It may be interesting to note that the expression multiplying $K_S$ 
takes into account effects of orbital precession (because it is non-vanishing
if and only if the orbit precesses) and therefore represents physical phenomena
neglected by previous fits.

\section{Fitting procedure and results}

The fitting of the out-of-plane recoil is in principle complicated 
because in addition to the overall factors $K_2$, $K_3$, and $K_S$ (which
are the same for any mass ratio or spins), any particular set of runs 
with the same initial separation and mass ratio (which we will term a ``block")  has 
idiosyncratic values of $\Phi_1$ and $\Phi_2$ that, although not
fundamentally interesting, need to be fit to the data.  Therefore,
in the 17 blocks of data we fit, there are formally $3+2\times 17=37$
fit parameters.  In the eight data blocks for which $\alpha_1^{\perp}=0$, 
$\Phi_1$ never enters, and in the two for which $\alpha_1^{\perp}=\alpha_2^{\perp}$,
$\Phi_1=\Phi_2$.  Therefore the actual number of fitting parameters
is 27, but this is still large enough that a multiparameter fit
would be challenging.  Fortunately, each $(\Phi_1,\Phi_2)$ pair
only affects a single data block.  We can therefore speed up the
fitting, and incidentally concentrate on only the interesting
parameters, if we (1)~pick some values of $K_2$, $K_3$, and $K_S$
that apply to all data blocks, then (2)~for each data block, 
find the values of $\Phi_2$ and possibly $\Phi_1$ that optimize
the fit, repeating this using new values of $\Phi_1$ and $\Phi_2$
for each block.  This gives an overall fit for the assumed values
of $K_2$, $K_3$, and $K_S$, having optimized over the uninteresting
$\Phi$ parameters.  

The fit itself needs to be performed assuming uncertainties on
each of the numerical measurements of the kick.  Each such
calculation is computationally expensive and systematic errors
are usually difficult to quantify, hence we do not have enough
information to do a true fit.  As a substitute, we assume that
for each block of data, the uncertainty $\sigma$ in each kick is either equal to a fraction
(fixed for all blocks) of the maximum magnitude kick in the
block, or to a fixed fraction of the individual kick itself.  
The former may be justified because some sources of error will
be independent of the phase of the angles when the holes merge,
but we note that the fit performed with the latter assumption
(that the uncertainty equals a fractional error of each kick)
yields very similar values for the fitting factors.  As we do not
know what the actual fractional error is, in either case we adjust it so that
for our best fit we get a reduced $\chi^2$ of roughly unity
(given our 95 data points and 27 fitting variables, this means
we need a total $\chi^2$ of about 68).  We then evaluate every
$(K_2,K_3,K_S)$ triplet using 
$\chi^2=\sum ({\rm pred-kick})^2/\sigma^2$.  

Minimizing $\chi^2$ as calculated with respect to the maximum kick per block,
we find that $\sigma^2=0.0005V^2_{\rm \parallel,max}({\rm block})$ gives 
$\chi^2/{\rm dof}= 1.0$. 
We note that this value for $\sigma$ is comparable to the numerical
error as measured by the difference in kicks computed at different resolutions, 
when available (e.g. for the new simulations presented here,
or the $q=0.25$ case presented in Table IV of \cite{Lousto:2008dn}).
It is also worth noting that for the data from the new
simulations, the uncertainty due to orbital 
precession, discussed in \S~\ref{sec:simulations}, is less than
$0.02V_{\rm \parallel,max}<\sigma$, and therefore cannot significantly 
affect the fit.
Our best fit is 
$K_2=30,540$~\kms, $K_3=115,800$~\kms, and
$K_S=17,560$~\kms, with $1\sigma$ ranges
$28,900-32,550$~\kms, $107,300-121,900$~\kms,
and $15,900-19,000$~\kms, respectively.  
We emphasize that all three coefficients
are indispensable in obtaining a good fit;
e.g. using the same definition of $\chi^2$ as above,
if $K_3=0$ then the best fit gives $\chi^2/{\rm dof}=2.0$, 
or if $K_S=0$ then the best fit gives $\chi^2/{\rm dof}=1.6$.

Minimizing $\chi^2$ as calculated with respect to each individual kick,
we obtain similar results.
In this case, $\sigma^2=0.0016V^2_{\rm \parallel}$ gives
$\chi^2/{\rm dof}= 1.0$.
Our best fit becomes
$K_2=32,092$~\kms, $K_3=108,897$~\kms, and
$K_S=15,375$~\kms, 
in agreement with the above fit to within $\sim 5\%$ for $K_2$ and $K_3$
and $\lsim 12\%$ for $K_S$.

In Table~\ref{table:data}
we compare the predicted out-of-plane kicks with the
measured ones for the entire data set.  
For fit \#1 minimizing the error with respect to the maximum kick per block,
of the 95 points,
only 4 agree to worse than 10\%, and all of those occur for
kicks with magnitudes much less than the maximum in their
data block.  Therefore these could represent small phase
errors rather than relatively large fractional velocity
errors.
Only 15 of the 95 points agree to worse than 5\%.
Fit \#2 minimizing the error with respect to individual kicks performs
even better in this regard, with only 10 points differing by more than 5\%
and only 2 points differing by more than 10\%.
In either case, the fits with the new ansatz
are significantly better than either the simpler $\eta^2$ fit
proposed by \cite{Lousto:2008dn} or the $\eta^3$ fit proposed 
by \cite{Baker:2008md}.

\begin{deluxetable}{c|r|r|r|r|r|r|r|r|r|r}
\tablecolumns{8}
\tablewidth{0pt}
\tablecaption{
Recoil data from \cite{Lousto:2008dn} (set A), \cite{Baker:2008md} (set B), this work (set C), \cite{Dain:2008ck} (set D), and \cite{Campanelli:2007cg} (set E).
The spin angles $\phi_1$ and $\phi_2$ are in radians, and the recoil velocities $V_{||}$ are in \kms.
The 8th column shows the result of a fit intended to minimize the error relative to the maximum recoil velocity per $(q,\alpha_1^{\perp},\alpha_2^{\perp})$ triplet
and the 9th column shows the relative error from that fit.
The 10th column shows the result of a fit intended to minimize the conventional relative error and the 11th column shows the relative
error from that fit.  In both cases, the vast majority of points agree to well within $10\%$.
}
\label{table:data}
\tablehead{\colhead{set} & \colhead{$q$} & \colhead{ $\alpha^{\perp}_1 $ } & 
\colhead{$\alpha^{\perp}_2 $} & \colhead{$ \phi_1$} & \colhead{$\phi_2 $} & 
\colhead{num. $V_{||} $} & \colhead{fit $V_{||}$ \#1} &
\colhead{$|\Delta V/V| $} & \colhead{fit $V_{||}$ \#2} & \colhead{$|\Delta V/V|$}   }
\startdata
A &0.125&0.000&0.751&0.000& 1.571& -113.3& -118.5& 0.046& -118.8& 0.049\\ 
A &0.125&0.000&0.756&0.000& 1.640& -101.9&  -95.9& 0.058&  -96.5& 0.053\\
A &0.125&0.000&0.761&0.000& 0.223& -349.1& -354.5& 0.016& -350.1& 0.003\\
A &0.125&0.000&0.747&0.000& 4.665&  131.6&  133.4& 0.014&  133.4& 0.014\\
A &0.125&0.000&0.772&0.000& 3.128&  338.7&  339.6& 0.003&  335.0& 0.011\\
A &0.167&0.000&0.777&0.000& 1.571&  530.6&  524.7& 0.011&  517.0& 0.026\\
A &0.167&0.000&0.739&0.000& 2.688&  348.8&  369.1& 0.058&  368.6& 0.057\\
A &0.167&0.000&0.786&0.000& 0.947&  320.3&  327.1& 0.021&  318.9& 0.004\\
A &0.167&0.000&0.773&0.000&-1.509& -529.0& -532.0& 0.006& -524.6& 0.008\\
A &0.167&0.000&0.778&0.000& 3.721& -130.8& -140.4& 0.074& -133.5& 0.021\\
A &0.250&0.000&0.779&0.000& 1.571& -909.9& -908.1& 0.002& -898.0& 0.013\\
A &0.250&0.000&0.788&0.000& 1.203& -815.9& -811.8& 0.005& -801.1& 0.018\\
A &0.250&0.000&0.760&0.000& 0.080&   56.3&   52.2& 0.074&   56.1& 0.004\\
A &0.250&0.000&0.787&0.000& 4.463&  866.0&  858.4& 0.009&  847.7& 0.021\\
A &0.250&0.000&0.751&0.000& 3.041& -209.9& -209.2& 0.003& -211.3& 0.007\\
A &0.333&0.000&0.794&0.000& 1.571&-1145.2&-1123.8& 0.019&-1109.6& 0.031\\
A &0.333&0.000&0.794&0.000& 1.164& -832.5& -833.6& 0.001& -819.3& 0.016\\
A &0.333&0.000&0.754&0.000& 0.082&  397.7&  387.9& 0.025&  391.8& 0.015\\
A &0.333&0.000&0.795&0.000& 4.466&  981.7&  968.3& 0.014&  953.8& 0.028\\
A &0.333&0.000&0.751&0.000& 3.017& -611.4& -603.8& 0.012& -604.9& 0.011\\
A &0.400&0.000&0.793&0.000& 1.571&-1414.7&-1399.6& 0.011&-1386.6& 0.020\\
A &0.400&0.000&0.798&0.000& 1.054&-1180.6&-1170.4& 0.009&-1155.3& 0.021\\
A &0.400&0.000&0.767&0.000& 0.017&   91.7&   83.9& 0.084&   91.3& 0.004\\
A &0.400&0.000&0.798&0.000& 4.426& 1338.6& 1319.5& 0.014& 1304.8& 0.025\\
A &0.400&0.000&0.760&0.000& 2.980& -328.3& -321.0& 0.022& -326.0& 0.007\\
A &0.500&0.000&0.771&0.000& 1.571&   34.0&   26.6& 0.217&   34.0& 0.000\\
A &0.500&0.000&0.777&0.000& 0.642& 1261.7& 1249.7& 0.010& 1244.4& 0.014\\
A &0.500&0.000&0.800&0.000&-0.323& 1528.2& 1493.9& 0.022& 1479.7& 0.032\\
A &0.500&0.000&0.764&0.000& 4.320& -622.6& -603.4& 0.031& -605.6& 0.027\\
A &0.500&0.000&0.800&0.000& 2.674&-1439.4&-1401.8& 0.026&-1387.2& 0.036\\
A &0.666&0.000&0.772&0.000& 1.571&  895.2&  872.5& 0.025&  865.9& 0.033\\
A &0.666&0.000&0.802&0.000& 0.506& 1699.3& 1672.6& 0.016& 1662.6& 0.022\\
A &0.666&0.000&0.798&0.000&-0.356& 1025.0& 1000.5& 0.024&  995.7& 0.029\\
A &0.666&0.000&0.784&0.000& 4.325&-1362.7&-1339.9& 0.017&-1330.8& 0.023\\
A &0.666&0.000&0.795&0.000& 2.654& -823.9& -813.3& 0.013& -809.8& 0.017\\
A &1.000&0.000&0.800&0.000& 1.571& 1422.3& 1422.6& 0.000& 1421.0& 0.001\\
A &1.000&0.000&0.803&0.000& 0.428& 1009.3&  990.5& 0.019&  980.9& 0.028\\
A &1.000&0.000&0.785&0.000&-0.461& -246.0& -237.4& 0.035& -245.3& 0.003\\
A &1.000&0.000&0.805&0.000& 4.222&-1479.9&-1468.0& 0.008&-1461.9& 0.012\\
A &1.000&0.000&0.786&0.000& 2.581&  390.5&  380.5& 0.026&  387.8& 0.007\\
B &0.333&0.200&0.022&0.000& 3.142&   49.0&   50.5& 0.031&   52.0& 0.062\\
B &0.333&0.200&0.022&5.498& 2.356&   48.0&   48.1& 0.003&   48.9& 0.018\\
B &0.333&0.200&0.022&4.712& 1.571&   17.0&   17.6& 0.034&   17.2& 0.012\\
B &0.333&0.200&0.022&0.000& 0.000&  114.0&  114.3& 0.003&  115.2& 0.011\\
B &0.500&0.200&0.050&0.000& 3.142&  -37.0&  -38.2& 0.033&  -36.7& 0.008\\
B &0.500&0.200&0.050&5.498& 2.356&  111.0&  111.2& 0.002&  114.1& 0.028\\
B &0.500&0.200&0.050&4.712& 1.571&  193.0&  195.5& 0.013&  198.0& 0.026\\
B &0.500&0.200&0.050&5.498& 1.571&   75.0&   73.2& 0.023&   76.4& 0.019\\
B &0.500&0.200&0.050&0.000& 1.571&  -55.0&  -56.7& 0.031&  -55.0& 0.000\\
B &0.666&0.200&0.089&1.047& 4.189& -381.0& -382.9& 0.005& -384.5& 0.009\\
B &0.666&0.200&0.089&0.000& 3.142& -135.0& -132.1& 0.022& -133.4& 0.012\\
B &0.666&0.200&0.089&5.498& 2.356&  168.0&  165.3& 0.016&  165.2& 0.017\\
B &0.666&0.200&0.089&4.712& 1.571&  364.0&  365.9& 0.005&  367.0& 0.008\\
B &0.769&0.200&0.118&0.000& 3.142& -386.0& -387.2& 0.003& -388.1& 0.005\\
B &0.769&0.200&0.118&5.498& 2.356& -525.0& -521.7& 0.006& -521.2& 0.007\\
B &0.769&0.200&0.118&4.712& 1.571& -348.0& -350.6& 0.008& -349.0& 0.003\\
B &0.909&0.200&0.165&0.000& 3.142& -542.0& -542.8& 0.001& -542.4& 0.001\\
B &0.909&0.200&0.165&5.498& 2.356& -657.0& -656.3& 0.001& -655.2& 0.003\\
B &0.909&0.200&0.165&4.712& 1.571& -384.0& -385.3& 0.003& -384.1& 0.000\\
C &0.674&0.367&0.236&2.513& 5.655& -973.7& -944.4& 0.030& -941.0& 0.034\\
C &0.674&0.367&0.236&2.094& 4.189& -442.9& -435.8& 0.016& -431.8& 0.025\\
C &0.674&0.367&0.236&2.094& 0.000& -988.5& -934.6& 0.055& -934.4& 0.055\\
C &0.674&0.367&0.236&1.257& 4.398& -361.7& -344.3& 0.048& -339.4& 0.062\\
C &0.674&0.367&0.236&0.000& 4.189&  374.2&  331.6& 0.114&  336.5& 0.101\\
C &0.674&0.367&0.236&0.000& 3.142&  749.0&  731.6& 0.023&  731.2& 0.024\\
C &0.674&0.367&0.236&0.000& 2.094&  672.4&  682.2& 0.015&  677.3& 0.007\\
C &0.674&0.367&0.236&5.027& 1.885&  826.1&  796.5& 0.036&  791.4& 0.042\\
C &0.674&0.367&0.236&4.189& 2.094&  619.7&  603.0& 0.027&  597.9& 0.035\\
C &0.674&0.367&0.236&4.189& 0.000& -245.6& -246.4& 0.003& -245.5& 0.000\\
C &0.674&0.367&0.236&3.770& 0.628& -237.7& -239.4& 0.007& -242.1& 0.019\\
C &0.515&0.367&0.177&2.513& 5.655& -584.3& -578.0& 0.011& -587.0& 0.005\\
C &0.515&0.367&0.177&2.094& 4.189& -123.0& -103.3& 0.160& -118.1& 0.040\\
C &0.515&0.367&0.177&2.094& 0.000& -684.8& -683.1& 0.002& -685.0& 0.000\\
C &0.515&0.367&0.177&1.257& 4.398&  -16.9&    1.0& 1.059&  -16.5& 0.024\\
C &0.515&0.367&0.177&0.000& 4.189&  483.9&  462.6& 0.044&  448.5& 0.073\\
C &0.515&0.367&0.177&0.000& 3.142&  579.8&  578.6& 0.002&  576.8& 0.005\\
C &0.515&0.367&0.177&0.000& 2.094&  345.2&  346.7& 0.004&  357.5& 0.036\\
C &0.515&0.367&0.177&5.027& 1.885&  362.4&  356.6& 0.016&  373.0& 0.030\\
C &0.515&0.367&0.177&4.189& 2.094&  202.1&  220.5& 0.091&  236.6& 0.171\\
C &0.515&0.367&0.177&4.189& 0.000& -231.1& -243.4& 0.053& -239.4& 0.036\\
C &0.515&0.367&0.177&3.770& 0.628& -354.8& -358.2& 0.010& -346.3& 0.024\\
D &1.000&0.930&0.930&1.571& 4.712& 2372.0& 2413.6& 0.018& 2423.7& 0.022\\
D &1.000&0.930&0.930&1.833& 4.974& 2887.0& 2972.2& 0.030& 2975.8& 0.031\\
D &1.000&0.930&0.930&2.269& 5.411& 3254.0& 3440.6& 0.057& 3432.8& 0.055\\
D &1.000&0.930&0.930&3.176& 0.035& 2226.0& 2390.4& 0.074& 2366.0& 0.063\\
D &1.000&0.930&0.930&4.817& 1.676&-2563.0&-2659.2& 0.038&-2666.8& 0.041\\
D &1.000&0.930&0.930&4.974& 1.833&-2873.0&-2972.2& 0.035&-2975.8& 0.036\\
D &1.000&0.930&0.930&5.149& 2.007&-3193.0&-3233.9& 0.013&-3232.9& 0.013\\
D &1.000&0.930&0.930&5.934& 2.793&-2910.0&-3152.3& 0.083&-3133.1& 0.077\\
E &1.000&0.515&0.515&1.571& 4.712& 1833.0& 1881.1& 0.026& 1875.6& 0.023\\
E &1.000&0.515&0.515&0.785& 3.927& 1093.0& 1079.6& 0.012& 1076.4& 0.015\\
E &1.000&0.515&0.515&3.142& 0.000&  352.0&  354.4& 0.007&  353.3& 0.004\\
E &1.000&0.515&0.515&4.712& 1.571&-1834.0&-1881.1& 0.026&-1875.6& 0.023\\
E &1.000&0.515&0.515&3.304& 0.162&   47.0&   46.3& 0.014&   46.2& 0.017\\
E &1.000&0.515&0.515&0.000& 3.142& -351.0& -354.4& 0.010& -353.3& 0.007 
\enddata
\end{deluxetable}

\section{Ejection probabilities and discussion}

One of the most important outputs of kick calculations and fits is 
the probability distribution of kicks given assumptions about
the mass ratio, spin magnitudes, and spin directions.  This
distribution is critical to studies of hierarchical merging in
the early universe (e.g., \citet{Volonteri:2007et}) as well 
as to the gas within galaxies \citep{Devecchi:2008qy} and an evaluation of the
prospects for growth of intermediate-mass black holes in globular
clusters \citep{HolleyBockelmann:2007eh}.  In Table~2
we show the results of our work (from fit \#1), compared with the proposed fit
formula of \cite{Campanelli:2007ew}.  It is clear that our work
gives distributions very close to those of \cite{Campanelli:2007ew},
with perhaps slightly smaller kicks because of the $\eta^3$
term we include.  

\begin{deluxetable}{lrrr}
\tabletypesize{\small}
\tablecolumns{4}
\tablewidth{0pt}
\tablecaption{Fraction of kick speeds above a given threshold, compared
with the results of \cite{Campanelli:2007ew}
(CLZM). In all cases we assume
an isotropic distribution of spin orientations.\label{tab:kickcomp}}
\tablehead{\colhead{Mass ratio and spin} & \colhead{Speed threshold} 
& \colhead{CLZM} &\colhead{This work}}
\startdata
$1/10 \leq q \leq 1$, $\alpha_1=\alpha_2=0.9$        & $v>500 \kms$  & 
0.364$\pm$0.0048  & 0.342526$\pm$0.00019\\
                                             & $v>1000 \kms$  
& 0.127$\pm$0.0034  & 0.120974$\pm$0.00011 \\

$1/4 \leq q \leq 1$, $\alpha_1=\alpha_2=0.9$         & $v>500 \kms$ & $
0.699 \pm 0.0045$  & 0.697818$\pm$0.00026\\
                                             & $v>1000 \kms$ & 
0.364$\pm$0.0046  & 0.353393$\pm$0.00019 \\

$1/4\leq q \leq 1$, $0\leq \alpha_1,\alpha_2\leq 1$         &  $v>500 \kms$ & 
 0.428$\pm$0.0045  & 0.415915$\pm$0.00020\\
&  $v>1000 \kms$ &
 0.142$\pm$0.0034  & 0.134615$\pm$0.00012\\
\enddata
\end{deluxetable}

In summary, we have demonstrated that a modified formula fits
all available out-of-plane kicks extremely well.  The wide range
of mass ratios, spin magnitudes, and angles explores all the
major aspects of parameter space for the out-of-plane kicks,
and thus we do not expect new results to deviate significantly
from our formula.  The excellence of these fits 
suggests that the kick distribution
is known to an accuracy that is sufficient for any
astrophysical purpose.

\acknowledgments

New simulations used for this work were performed on Jaguar at Oak Ridge National Laboratory. 
MCM acknowledges partial support from the National Science Foundation under grant AST 06-07428 and NASA ATP grant NNX08AH29G.  
The work at Goddard was supported in part by NASA grant 06-BEFS06-19.
We also wish to think S. McWilliams and A. Buonanno for helpful discussions.

\clearpage

\bibliography{refs}

\end{document}